\documentclass{aa}
\usepackage{graphicx}
\usepackage{amsmath,amssymb}
\usepackage{txfonts}
\usepackage{color,ulem}
\newcommand{\dd}{\mathrm{d}}
\newcommand{\rg}{r_{\mathrm{g}}}
\newcommand{\rp}{r_{\mathrm{p}}}
\newcommand{\ra}{r_{\mathrm{a}}}
\newcommand{\rorb}{a}
\newcommand{\MBH}{M_{\mathrm{BH}}}
\newcommand{\Mdisk}{M_{\mathrm{disk}}}
\newcommand{\Menc}{M_{\mathrm{enc}}}
\newcommand{\DO}{\Delta\Omega}
\newcommand{\Dw}{\Delta\omega}

\providecommand{\fdg}{\hbox{$.\!\!^\circ$}}
\begin{document}

\title{S301 and friends: Measuring the spin of Sgr~A*}
\titlerunning{S301 and friends}
\author{T. Piran\inst{1}
\and P. Amaro-Seoane\inst{2,3}
\and B. Aytac\inst{3}
\and G. Bourdarot\inst{3}
\and A. Burkert\inst{4,3}
\and D. Calder{\'o}n\inst{5}
\and J. Cuadra\inst{6,7,3}
\and F. Eisenhauer\inst{3,8}
\and R. Genzel\inst{3,9}
\and S. Gillessen\inst{3}
\and S. Joharle\inst{3}
\and F. Mang\inst{3,8}
\and T. Ott\inst{3}
\and D.C. Ribeiro\inst{3}
\and M. Sadun Bordoni\inst{3}
\and R. Sari\inst{1}
\and F. Thiel\inst{3}}
\authorrunning{Piran et al.}
\institute{Racah Institute of Physics, The Hebrew University, Jerusalem 91904, Israel\\
\email{tsvi.piran@mail.huji.ac.il}
\and Universitat Polit\`ecnica de Val\`encia, C/Vera s/n, Val\`encia, 46022, Spain
\and Max Planck Institute for Extraterrestrial Physics, Giessenbachstra{\ss}e 1, 85748 Garching, Germany
\and University Observatory, Faculty of Physics, Ludwig-Maximilians-Universit\"at, Scheinerstra{\ss}e 1, 81679 Munich, Germany
\and Max Planck Institute for Astrophysics, Karl-Schwarzschild-Stra{\ss}e 1, 85748 Garching, Germany
\and Universidad Adolfo Ib{\'a}\~nez, Av. Padre Hurtado 750, Vi\~na del Mar, Chile
\and Millennium Nucleus on Transversal Research and Technology to Explore Supermassive Black Holes (TITANS), Chile
\and 
Technical University of Munich, TUM School of Natural Sciences, Physics Department, 85747 Garching, Germany
\and Departments of Physics \& Astronomy, Le Conte Hall, University of California, Berkeley, CA 94720, USA
}
\date{Received ...; accepted ...}
\abstract
{
The discovery of S301, with pericenter distance $\rp\simeq 280\rg$ and eccentricity $e\simeq0.9825$, raises the prospect of measuring the spin parameter, $\chi$,
of Sgr A* through Lense-Thirring (LT) nodal precession. A major obstacle is Newtonian confusion: any nonspherical extended mass distribution can also induce nodal precession.}
{We aim to separate the LT spin signal of S301 from the nodal precession induced by an extended Newtonian mass distribution.}
{We compared the secular Newtonian torque exerted by a disk or flattened mass distribution on the orbits of S301 and of the apocenter-matched reference stars S2, S55, and S38 using analytic estimates validated by numerical orbit-averaged torque calculations.}
{For a disk or flattened distribution extending beyond the stellar pericenters, the secular Newtonian torque on a highly eccentric orbit is controlled mainly by the apocenter, whereas the LT signal is controlled mainly by the pericenter. Thus, stars with apocenters comparable to S301's but much larger pericenters, in particular S2 but also S55 and S38, experience comparable Newtonian torques while having $\sim 30$ times smaller LT signals (for S2). Their measured precessions, or the upper limits on them, can therefore be used to calibrate the mass and orientation of the Newtonian background for subtraction from S301's precession. The Schwarzschild apsidal advance of S301 further rotates the orbit's pericenter relative to any disk, producing a systematic time dependence in the Newtonian contribution, while the LT signal remains fixed by the spin vector. The granularity of the perturber population sets a stochastic floor on this subtraction, which orbit- and star-averaging suppress.}
{With continued GRAVITY+ astrometry and Extremely Large Telescope spectroscopy, the in-plane spin projection may be within reach in the near term; the full spin vector requires a much longer-term accumulation of the LT apsidal signal.}

\maketitle
\nolinenumbers
\section{Introduction}
\label{sec:intro}
Sgr~A* is the only black hole for which both the magnitude and the direction of its spin can be measured directly. 
{The spin is a fundamental parameter of the Kerr metric: measuring it enables tests of general relativity in the strong field, including, eventually, the no-hair relation between spin and quadrupole \citep{will2008}.  It also encodes the black hole's growth history, with a high spin indicating coherent accretion and a low spin chaotic accretion or mergers \citep[e.g.,][]{Berti2008}, and it is a key input for models of Sgr A*'s accretion flow and its Event Horizon Telescope (EHT) image \citep{eht2022}.}

This is made possible by precise monitoring of stellar orbits in the Galactic Center.
A spinning black hole drags the
spacetime around it, and this frame dragging, the Lense-Thirring (LT)
effect \citep{lense1918}, makes each stellar orbit precess: the orbital
plane drifts (nodal precession, $\DO_{\mathrm{LT}}$) and so does the
pericenter direction (apsidal precession, $\Dw_{\mathrm{LT}}$).
Both {per orbit} rates grow steeply toward the black hole. They scale as $(\rg/p)^{3/2}$, {where $p=a(1-e^2) \approx  2 \rp $ for the highly eccentric orbits of interest,
$\rp$ is the pericenter distance, and $\rg\equiv G \MBH/c^2 $ }is the gravitational radius of
Sgr~A*. We focused on the nodal precession since the apsidal LT signal is masked by
the far larger Schwarzschild advance (see Sect.~\ref{sec:apsidal}).

The pericenter of S2, which had until recently been the nearest known star to Sgr~A*, is
$\rp^{\mathrm{S2}} = 2800\,\rg$ \citep{gillessen2017,gravity2020}. Its LT nodal rate (even for maximal spin and favorable orientation) is
$\DO_{\mathrm{LT}}^{\mathrm{S2}} = 3.3\times10^{-5}$~rad~orbit$^{-1}$,
too small for current instrumentation. Measuring the spin requires a star that passes about a factor of 3 closer to Sgr~A* \citep{Waisberg2018}.

\begin{figure*}[t]
\centering
\includegraphics[width=0.95\textwidth]{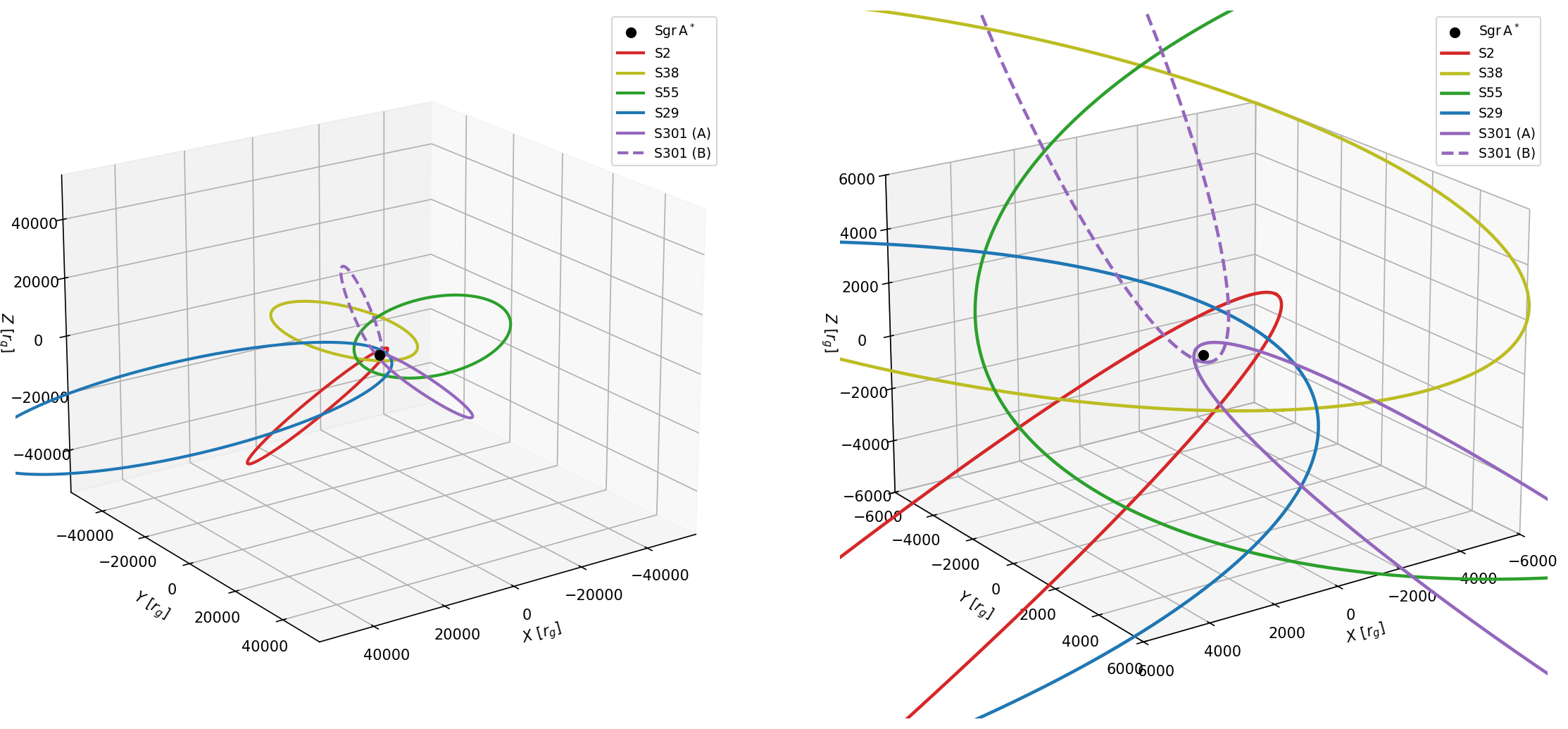}
\caption{Orbits of the benchmark S-stars and S301 around Sgr~A* in
gravitational radii, projected in three dimensions. \textit{Left}: Full-scale view; the
apocenters set the scale.
\textit{Right}: Zoom into the inner $\sim 6000\,\rg$, showing the disparate
pericenters and the strongly inclined, mutually misaligned orbital planes.
S301 is shown for its two degenerate orbit solutions (A and B). The inner
tracers S301, S55, S2, and S38 have comparable apocenters
($31{,}700$-$50{,}400\,\rg$) but pericenters that differ by more than an
order of magnitude. Their distinct orientations provide the
geometric diversity needed to resolve the spin vector.}
\label{fig:orbits}
\end{figure*}

The newly discovered S301 \citep{abdeldayem2026}, with $\rp^{\mathrm{S301}} =
280\,\rg$ and period $P = 8.7$~yr, is precisely such a star. For favorable spin-orbit orientation and maximal spin,  $\DO_{\mathrm{LT}}^{\mathrm{S301}}
= 9.6\times10^{-4}$~rad~orbit$^{-1}$ is some 30 times larger than $\DO_{\mathrm{LT}}^{\mathrm{S2}}$, 
making  spin
detection feasible within a few
orbital periods of continued GRAVITY+ monitoring when combined with future Extremely Large Telescope (ELT)
spectroscopy. Indeed, a mock-data analysis in the discovery paper — assuming $100\mu as$ GRAVITY+ astrometry and 1 km s$^{-1}$ ELT/MICADO radial velocities through 2035 but a pure Kerr metric with no extended mass — forecasts $\sigma_\chi < 0.2$ \citep{abdeldayem2026}.  
The current astrometric solution for S301 is not yet
unique: the data admit two nearly degenerate orbits, labeled A and B in
Fig.~\ref{fig:orbits}, which differ chiefly in orientation but have
comparable pericenters and apocenters.

A potential obstacle is the Newtonian confusion: any nonspherical mass
distribution produces nodal precession of the same character as LT.
\citet{RubilarEckart2001} first showed that for Galactic center stars
the Newtonian mass precession can rival, and even dominate, the
relativistic apsidal advance.
\citet{merritt2010} showed that a flattened stellar cusp of enclosed mass
{$\Menc$} {within the stellar orbit   (see Sect.~\ref{sec:disk})} 
produces nodal precession comparable to LT when
{{$\Menc/\MBH \sim (\rg/p)^{3/2}$}; for moderate eccentricities this is the familiar $(\rg/a)^{3/2}$ criterion, while for the highly eccentric orbits considered here it is set by the pericenter $(\rg/2 \rp)^{3/2}$}. \citet{iorio2011}
computed specific disk configurations and confirmed the degeneracy.
\citet{will2008} noted that using multiple stars with different $\rp$
helps break degeneracies since LT scales more steeply with $\rp$ than
Newtonian perturbations. Similarly, \citet{Heissel2022} showed that
within a single orbit of S2, the Newtonian mass precession is sourced
almost entirely in the apocenter half of the orbit, whereas the
relativistic precession arises near pericenter, enabling their
separation. The apocenter dominance of the secular Newtonian torque
that we derive below (Sect.~\ref{sec:disk}) is the multi-star, nodal
counterpart of this behavior.

Here we highlight a property specific to the S2+S301 pair: both stars
have comparable apocenters ($r_{\mathrm{a}}^{\mathrm{S301}} \approx
31700\,\rg$ and $r_{\mathrm{a}}^{\mathrm{S2}} \approx 45400\,\rg$): they differ from each other by less a factor of $\sim 1.4$.
This suggests  a joint S2+S301 strategy that can be used to break the degeneracy.
The same holds for S55 and S38, whose apocenters likewise lie within a
factor of $\sim1.6$ of S301's (Table~\ref{tab:combined_apocenters_rg}).
Throughout this paper we therefore compare S301 with S2 as a concrete,
best-measured representative of the apocenter-matched reference stars; the
same analysis applies, with minor numerical changes, to S55 and S38.

The underlying rationale can be summarized as follows: (i) The LT precession scales as $(\rg/\rp)^{3/2}$, so S301's signal exceeds S2's by a factor of $\sim30$.
(ii) For any disk or flattened distribution extending beyond the stellar pericenters, the secular Newtonian torque is set by the apocenter (which we derive in Sect. \ref{sec:disk} and validate numerically in Sect. \ref{sec:numerical}); S301, S2, S55, and S38 have comparable apocenters, so they feel comparable Newtonian torques. (iii) The wide-pericenter stars therefore calibrate the Newtonian background, which can then be subtracted from S301's precession, isolating the spin signal.

A single reference star is insufficient for this strategy: the disk-induced
nodal precession is a three-dimensional effect that depends on each orbit's
inclination and node relative to the unknown disk plane. One star therefore
constrains only one combination of the disk's mass and orientation. The
apocenter-matched reference stars S2, S55, and S38, whose orbital planes are
strongly mutually misaligned (Fig.~\ref{fig:orbits}), together triangulate
the disk's orientation as well as its mass, while the wider orbit of S29
anchors the radial profile -- all four are needed. Future detections of other stars with comparable apocenters will, of course, sharpen the ability to { pinpoint}  the geometry. 

Any mass distribution around Sgr~A* {extending over the region probed by the stellar orbits} 
is constrained by
the apsidal precession of S2 \citep{gravity2020,gravity2024}, with the
most recent analysis giving $\Menc \lesssim 1200\,M_\odot$ \citep{gravity2024}. {More compact configurations, buried inside the stellar pericenters, are ruled out separately.} Among all axisymmetric mass distributions with a given
radial profile, a razor-thin disk maximizes the quadrupole moment and
hence produces the largest Newtonian nodal precession; more spheroidal
distributions, such as ellipsoidal  configurations, produce
weaker nodal precession for the same enclosed mass.   Our numerical results
for disk configurations therefore represent reasonable upper limits on the Newtonian
confusion, {i.e., the possibility that nodal precession induced by a nonspherical mass distribution mimics the frame-dragging signal}  for a given mass. We therefore evaluated the Newtonian confusion with 
{a razor-thin disk normalized such that its mass within the stellar apocenters equals the enclosed mass, $\Menc$ (for our fiducial truncation $r_{\rm disk }= r_{\rm a}^{S301}$, this is the total disk mass, i.e., $\Mdisk = \Menc$).}
We adopted $\Menc = 10^3\,M_\odot$ as our fiducial value and a maximal black
hole spin $\chi = 1$ throughout. Since the nodal precession per orbit $\DO_{\mathrm{LT}} \propto \chi$,
smaller spin values are proportionally harder to detect.

 The paper is structured as follows. We derive in Sect.~\ref{sec:disk}
the apocenter-dominance of the Newtonian nodal precession.
Section~\ref{sec:LT_contrast} presents the LT precession and the
disk-to-LT confusion ratio. We  provide
numerical validation of the analytic estimates in Sect.~\ref{sec:numerical}. In Sect.~\ref{sec:joint_strategy}
we develop the joint strategy for separating disk and spin
signals. We briefly discuss the apsidal
precession in Sect.~\ref{sec:apsidal}. Motivated by \citet{Bordoni+2025}, we discuss the potential nuisance effect of granularity in Sect.~\ref{sec:granularity}, and we  introduce the time-variation
discriminant from Schwarzschild apsidal advance in Sect.~\ref{sec:time}. We conclude in
Sect.~\ref{sec:conclusions}.

\section{Disk torques}
\label{sec:disk}

\begin{figure*}[t]
\centering
\includegraphics[width=0.95\textwidth]{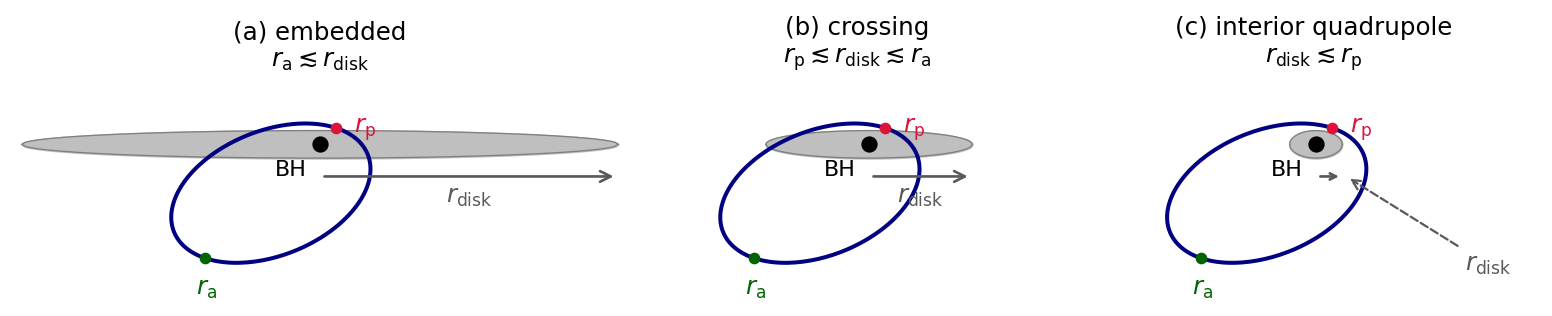}
\caption{Three orbit--disk configurations, shown edge-on for the same
inclined eccentric orbit (drawn with $e=0.75$ and $i=35^\circ$ for clarity;
the black circle marks the black hole, red and green dots the pericenter
and apocenter): (a) embedded, $r_{\rm a}\lesssim r_{\rm disk}$; (b)
crossing, $r_{\rm p}\lesssim r_{\rm disk}\lesssim r_{\rm a}$; (c) interior
quadrupole, $r_{\rm disk}\lesssim r_{\rm p}$. In (a) and (b) the secular
nodal torque is dominated by the apocenter, while in (c) it is set by the
semi-latus rectum (Eq.~\ref{eq:exterior}). Cases (a) and (b) are the
relevant ones: any disk extending beyond S2's pericenter places both stars
in the embedded or crossing regime, while (c) requires a fine-tuned mass
configuration.}
\label{fig:regimes}
\end{figure*}

We considered an axisymmetric disk with surface
density $\Sigma(r) \propto r^\alpha$, total mass $\Mdisk$, and outer
radius $r_{\mathrm{disk}}$. We calculated the corresponding secular nodal precession   using the Gauss planetary equation
\citep{BrouwerClemence1961,MurrayDermott1999}:
\begin{equation}
  \frac{\dd\Omega}{\dd t}
  = -\frac{1}{h\sin i}
    \frac{\partial\langle\Phi_{\mathrm{disk}}\rangle}{\partial i},
  \label{eq:dOdt}
\end{equation}
where $\Omega$ is the longitude of the ascending node, $i$ the inclination, and $\omega$ the argument of pericenter, all measured relative to the disk plane (the transcription to the observable sky-plane elements is discussed in Sect.~\ref{sec:LT_contrast}); $h = \sqrt{G\MBH\,\rorb(1-e^2)}$ is the specific angular momentum, and
$\langle\Phi_{\mathrm{disk}}\rangle$ is the disk potential averaged over
one orbital period\footnote{The averaging is done considering the orbit in the background field of the supermassive black hole, so it 
{assumes } that $\Mdisk\ll \MBH $.}, evaluated in true anomaly $f$ with the Kepler time
weight $\dd t \propto r^{2}\,\dd f$ (Eq.~\ref{eq:inv_delta}).

The secular disk torque is {controlled by the apocenter region, }
$r\sim r_{\rm a}=a(1+e)$ rather than the pericenter, provided the
disk extends beyond the stellar pericenter, $r_{\rm disk}\gtrsim
r_{\rm p}$. Two effects combine to produce this. The perturbation
exerted by an exterior ring at $r'$ scales as $(r/r')^2$ and is
therefore strongest at the largest radii the orbit reaches, where the
star also spends most of its period. {Near pericenter the dominant contribution comes from the small enclosed mass, $M(<r) \propto  r^{\alpha+2}$, acting as an interior quadrupole, while the effect of exterior material is suppressed by $(r/r')^2$; the pericenter contribution to the secular precession is therefore small}.

When $r_{\rm a}\lesssim r_{\rm disk}$ (orbit embedded in the disk; Fig.~\ref{fig:regimes}a)
the torque is dominated by 
{material near the apocenter}
$r'\sim r_{\rm a}$, and the net precession per orbit 
{scales as}
\begin{equation}\label{eq:embedded}
\Delta\Omega_{\rm disk}\sim
\frac{M(< r_{\rm a})}{M_{\rm BH}}  
g(e,\alpha)\cos i,
\end{equation}
where $g(e,\alpha)$ is a dimensionless order-unity function 
{describing the eccentricity dependence at fixed enclosed mass; the dependence on the semimajor axis enters through $M(< r_{\rm a})$.}
The form of $g$
depends on $\alpha$: for $\alpha=0$ (uniform surface density) the
apocenter-dominated torque integral gives $g(e,0)\propto
1/[(1+e)^{3/2}(1-e)^{1/2}]$, which
diverges\footnote{For S301 ($1 - e \approx 0.018$) the
corresponding enhancement is a factor of $\simeq7.6$, and in practice the divergence is cut off once the pericenter approaches the disk's inner edge or when finite thickness of the disk is considered.} as $e\to 1$;
for
$\alpha=-1$ (Mestel disk, $\Sigma\propto 1/r$) the eccentricity
dependence nearly cancels and $g(e,-1)$ is approximately constant.

When $r_{\rm p}\lesssim r_{\rm disk}\lesssim r_{\rm a}$ (crossing
orbit; Fig.~\ref{fig:regimes}b) the star leaves the disk on its way to apocenter. The
pericenter contribution remains suppressed by the small enclosed
mass, and the torque is dominated by material near the disk's outer
edge and by the disk-plane crossings.
The
precession is again $\Delta\Omega_{\rm disk}\sim(M_{\rm disk}/M_{\rm
BH})\,\tilde g(e,\alpha,r_{\rm a}/r_{\rm disk})\cos i$, joining
smoothly onto Eq.~(\ref{eq:embedded}) at $r_{\rm a}\sim r_{\rm
disk}$; the pericenter enters only weakly, through the node-crossing
radii. In both regimes the precession is set by the apocenter and is
maximal when $r_{\rm a}\sim r_{\rm disk}$.

The picture changes qualitatively only when the entire orbit lies
outside the disk, $r_{\rm p}\gtrsim r_{\rm disk}$ (Fig.~\ref{fig:regimes}c). The disk then acts
as an interior quadrupole, $Q\sim M_{\rm disk}r_{\rm disk}^2$
(dominated by the outer edge for any $\alpha>-4$), and the standard secular result is
\begin{equation}\label{eq:exterior}
\Delta\Omega_{\rm disk}\simeq -3\pi\,k(\alpha)\,
\frac{M_{\rm disk}}{M_{\rm BH}}
\left(\frac{r_{\rm disk}}{p}\right)^{2}\cos i,
\end{equation}
with $p\equiv a(1-e^2)$ and $k(\alpha)=(\alpha+2)/[2(\alpha+4)]$. Although the star spends
most of its period near apocenter, the $r^{-3}$ torque weighting
dominates the time average, $\langle(a/r)^3\rangle
=(1-e^2)^{-3/2}$, so the precession {per orbit } is set by the semi-latus rectum
$p\approx2r_{\rm p}$ that is by the pericenter, not the apocenter.

One can compare now the disk torques on S301
and S2 (the latter representing the apocenter-matched reference stars) as a function of the disk size, from the largest disks (the
relevant case) down to the most compact.
Since S2 and S301 have comparable apocenters,
$r_{\rm a}^{\mathrm{S2}}\approx 4.54\times10^4\,r_{\rm g}$ and
$r_{\rm a}^{\mathrm{S301}}\approx 3.17\times10^4\,r_{\rm g}$,
a ratio of only $\sim1.4$, both stars avoid the interior-quadrupole
regime for any disk extending beyond S2's pericenter,
$r_{\rm disk}\gtrsim 2.8\times10^3\,r_{\rm g}$.
They therefore sample the same broad apocenter-dominated or apocenter crossing
regimes, and the ratio of their disk-induced nodal precession rates
is expected to remain of order unity. This conclusion is insensitive,
up to factors of order unity, to the detailed choice of $\alpha$,
$r_{\rm disk}$, and the functions $g$ or $\tilde g$ that describe the
embedded and crossing regimes.
In the transitional case where $r_{\mathrm{disk}}$ falls between the two
apocenters, applying one regime to each star gives
$\DO_{\mathrm{disk}}^{\mathrm{S2}}/\DO_{\mathrm{disk}}^{\mathrm{S301}}$  of order unity.

For an intermediate disk, $280\,r_{\rm g}\lesssim r_{\rm disk}\lesssim 2800\,r_{\rm g}$,
S301 still crosses the disk while S2's orbit lies entirely outside it: S2 then responds
only to the interior quadrupole (Eq.~\ref{eq:exterior}), suppressed by
$(r_{\rm disk}/r_{\rm p}^{\mathrm{S2}})^2$, and the torque ratio drops below unity. In this case S2
underestimates the disk torque on S301, and the separation must rely instead on the
time-variation discriminant (Sect.~\ref{sec:time}), which operates only on much longer
timescales. However,  this scenario requires a {very compact configuration that is excluded as discussed below }. Moreover,  the enclosed mass $M_{\rm enc}$, constrained only to lie within S2's
orbit, would have to be concentrated inside a fraction of S2's pericenter, a radius $\sim 16$ times
smaller than its apocenter.

Both orbits lie entirely outside the disk it the disk is compact enough to fit inside S301's own pericenter,
$r_{\rm disk}<280\,r_{\rm g}$. In this case Eq.~(\ref{eq:exterior}) applies to both stars. Because
$\Delta\Omega_{\rm disk}$ in this regime scales as $r_{\rm disk}^{2}$,
its effect on S301 falls off rapidly for a disk well inside the
pericenter; we quantify this in Sect.~\ref{sec:LT_contrast}
(Eq.~\ref{eq:compact_ratio}). Such a compact disk is therefore not a
significant contaminant. Moreover, a configuration in which all the mass
interior to S2's apocenter is compressed into a disk smaller than S301's
pericenter (a radius $\sim160$ times smaller) and at the same time has a significant quadrupole moment, is even more fine-tuned than the intermediate case above.

The data already bound the Newtonian torque on S2 directly.
Detecting the LT precession would require a star whose pericenter is at
least three times smaller than $r_{\rm p}^{\rm S2}$
\citep{abdeldayem2026}; the corresponding non-detection for S2 bounds its
total nodal precession to
$\lesssim 3^{3/2}\,\Delta\Omega^{\rm S2}_{\rm LT}\approx
1.7\times10^{-4}$~rad~orbit$^{-1}$. For a distribution buried inside
S2's pericenter this translates, via Eq. \ref{eq:exterior}, into a bound on the quadrupole,
$Q\cos i_{\rm S2} \sim M_{\rm disk}\,r_{\rm disk}^{2}\cos i_{\rm S2}
\lesssim 10^{10}\,M_\odot\,r_{\rm g}^{2}$, where $i_{\rm S2}$ is the
inclination of S2's orbit with respect to the disk plane\footnote{{A disk
oriented nearly perpendicular to S2's orbit ($\cos i_{\rm S2}\to0$)
evades the nodal bound, but is independently constrained at a
comparable level by S2's measured apsidal precession, whose quadrupole
contribution, $\propto 2-{5\sin^{2}i}/{2}$, does not vanish for a
polar orbit; such an orientation would in any case add a further layer
of fine-tuning.}}.

{This bound constrains only the combination $M_{\rm disk}r_{\rm disk}^{2}$:
dynamically, it still permits masses as large as
$\sim\!10^{5}\,M_\odot$ for a disk truncated near S301's pericenter,
$r_{\rm disk}\approx r_{\rm p}^{\rm S301}=280\,r_{\rm g}$.
Two independent arguments exclude such a configuration.}

{First, it requires a fine-tuned outer edge. The cumulative quadrupole of a disk grows steeply outward, $Q(<r)\propto\int\Sigma r^{3}{\rm d}r
\propto r^{\alpha+4}$, so extending the same surface-density law by even
a factor of 3 in radius increases the S2-visible quadrupole by
$3^{\alpha+4}$ (a factor of 27 for $\alpha=-1$) and violates the bound
above. The disk must therefore be truncated within a factor of
$\sim\!2$--$3$ in radius above S301's pericenter. No physical scale --
innermost stable circular orbit, tidal-truncation radius, or resonance -- exists at
$\sim\!280\,r_{\rm g}$ to produce such an edge; the only feature that
singles out this radius is S301's orbit itself.  It is for this reason that we refer to such configurations as ``fine-tuned" in this paper.}

{Second, no known form of matter can supply the required
$\gtrsim\!10^{4}\,M_\odot$ within $280\,r_{\rm g}$ of Sgr~A*. The hot gas at these
radii is directly measured --- $n_{\rm e}\sim10^{6}\,{\rm cm^{-3}}$ at a
few $r_{\rm g}$ from EHT submillimeter modeling \citep{eht2022}, a few
$\times10^{3}\,{\rm cm^{-3}}$ at $\sim\!10^{3}\,r_{\rm g}$ from the drag
on G2 \citep{gillessen2019}, and $\sim\!10^{2}\,{\rm cm^{-3}}$ at the
Bondi radius \citep{baganoff2003,wang2013}, consistent with the
rotation-measure accretion rate of $\sim\!10^{-8}\,M_\odot\,{\rm yr^{-1}}$
\citep{marrone2007} and with radiatively inefficient accretion flow models  \citep{yuan2003,yuan2014}. Integrating any profile through these values
yields $\lesssim\!10^{-8}\,M_\odot$ inside $280\,r_{\rm g}$. }

{A cooler,
massive disk cannot persist: the implied surface density,
$\Sigma\sim10^{8}\,{\rm g\,cm^{-2}}$, requires $T\gtrsim10^{5}$~K for
Toomre stability. Any cooler disk fragments into stars within days
(the local orbital period is $\sim\!1$~week). A Toomre-stable disk
has a cooling time that  is a fraction of a second and its
emission  would exceed Sgr~A*'s observed
$\sim\!10^{-9}\,L_{\rm Edd}$ output by many orders of magnitude.}

{Main-sequence stars are excluded:  (i)  $10^{4}$ solar-type stars
would appear as an $m_{\rm K}\approx10$ source coincident with Sgr~A*
(versus the observed $m_{\rm K}\approx17$--$18$). (ii)  At the local
orbital velocity of $\sim\!0.06c$ their collision time is under a year,
each collision releasing $\sim 10^{51}$ erg.
Giants cannot exist there at all, their tidal radii
($\sim\!10^{3}\,r_{\rm g}$) exceeding the region. Planetesimals are
worse still: at fixed total mass the collision timescales as
$m^{1/3}$, giving Jupiter-mass bodies a collision time of weeks and
grinding the population into gas, which is excluded as above. }

{At such small distances from the black hole compact
remnants are removed by gravitational-wave inspiral
($t_{\rm GW}\approx3.5\times10^{7}$~yr for a $10\,M_\odot$ black hole on
a circular orbit at $280\,r_{\rm g}$ and faster for elliptical orbits with this apocenter). Sustaining $10^{4}\,M_\odot$
would require a steady inflow of $\sim\!3\times10^{-4}\,M_\odot\,{\rm
yr^{-1}}$ of compact objects, orders of magnitude beyond any
replenishment channel and independently capped by the drain limit
\citep{alexander2004}. 
A single body evades these arguments only as an
intermediate-mass black hole {of mass $10^4,M_\odot$,} for which $t_{\rm GW}\sim3\times10^{4}$~yr at this radius.
Finally, a component dark and collisionless enough to survive (e.g., a
dark-matter spike) is dissipationless and hence spherical, with no
quadrupole and no nodal precession.}

{In short, the contaminating configuration must be flattened; flattening
requires dissipation; dissipation requires baryons; and every baryonic
reservoir of $\gtrsim\!10^{4}\,M_\odot$ inside 280 $\rg$ is excluded. A
compact disk capable of mimicking the LT precession of S301 is
therefore ruled out on both dynamical and astrophysical grounds.}

\section{The Lense-Thirring precession}
\label{sec:LT_contrast}
\begin{figure}[t]
\centering
\includegraphics[width=0.8\hsize]{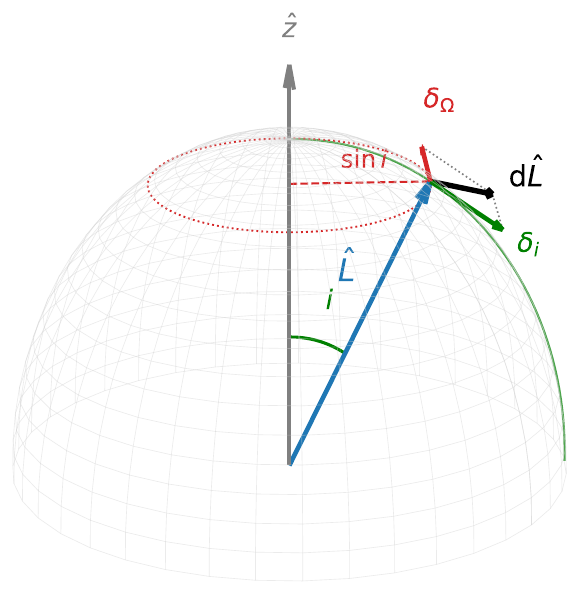}
\caption{Decomposition of the orbital-pole drift. Any torque rigidly tilts
the orbital plane, producing a frame-independent drift, $\mathrm{d}\hat{\mathbf{L}}$, of the orbital pole, $\hat{\mathbf{L}}$, which is  at colatitude $i$ from the reference-plane normal, $\hat{z}$. The drift
decomposes into a meridional part, $\delta_i$, which changes the inclination
directly ($\dd i=\delta_i$; Eq.~\ref{eq:DiLT}), and an azimuthal part,
$\delta_\Omega$, which moves the node along a circle of radius $\sin i$ such
that $\Delta\Omega=\delta_\Omega/\sin i$.
The nodal rate grows toward small $i$, and in the limit $i\to0$ the node is
undefined.}
\label{fig:decomp}
\end{figure}

The LT nodal rate per orbit is \citep[e.g.,][]{will2008}\begin{equation}
\label{eq:DOLT1}
  \DO_{\mathrm{LT}}
  = \frac{4\pi\chi\,{\sin\beta\sin\lambda}}{(1-e^2)^{3/2}\,\sin i}
    \left(\frac{\rg}{\rorb}\right)^{\!\!3/2} 
  = \frac{4\pi\chi\,{\sin\beta\sin\lambda}}
         {(1+e)^{3/2}\,\sin i}
    \left(\frac{\rg}{\rp}\right)^{\!\!3/2} \ .
\end{equation}
Here $i$ is the inclination of the orbit with respect to the reference
(disk) plane. The factor $1/\sin i$ is the kinematic projection of the rigid
precession of the orbital plane onto the motion of the node measured in that
plane (Fig.~\ref{fig:decomp}).
When the reference plane is the black hole's equatorial plane, $\sin\beta\sin\lambda=\sin i$ and the familiar
frame-independent rate is recovered.
$\beta$ is the angle between the spin of Sgr~A* and the orbital angular
momentum, $\mathbf{L}$, and $\lambda$ is the azimuth of the spin projection onto
the orbital plane, measured from the ascending node on the disk plane;
$\beta$ is frame-independent, whereas $\lambda$, like $\Omega$ and $\omega$,
depends on the reference plane through the node. 
With this convention the
components of the spin in the orbital frame are
$\chi(\sin\beta\cos\lambda,\ \sin\beta\sin\lambda,\ \cos\beta)$, and the three
LT rates measure them directly: the nodal and inclination drifts carry the two
in-plane components, while the 
{node-corrected apsidal combination obeys
$\Delta\widetilde\varpi_{\rm LT}\equiv\Dw_{\mathrm{LT}}+\cos i\,\DO_{\mathrm{LT}}
\propto\chi\cos\beta$ (Sect.~\ref{sec:apsidal})\footnote{The angle
$\widetilde\varpi$, whose drift is
$\dot{\widetilde\varpi}=\dot\omega+\cos i\,\dot\Omega$, differs from the
standard celestial-mechanics longitude of pericenter,
$\varpi\equiv\omega+\Omega$ \citep{MurrayDermott1999}, to which it reduces for
$i\to0$. The $\cos i$ factor cancels the kinematic nodal term in
$\dot\omega$ such that $\Delta\widetilde\varpi_{\rm LT}$ takes the same form
in any reference plane.}.}
The numerical rates quoted below
are the frame-invariant plane-precession magnitudes,
$4\pi\chi\sin\beta\,(1-e^2)^{-3/2}(\rg/\rorb)^{3/2}$ per orbit, evaluated for
the most favorable orientation $\sin\beta = 1$, consistent with the
maximal-spin convention $\chi = 1$; nodal rates referred to the disk plane are
larger by the kinematic factor $1/\sin i$.

All rates in Sects.~\ref{sec:disk}--\ref{sec:numerical} are referred
to the disk plane, where the Newtonian torque takes its simplest form. The
physical content of both perturbations is the vector precession,
$\dot{\hat{\mathbf{L}}}$, of the orbital pole;
projecting this vector onto the observable sky-plane element rates
involves only each star's measured $(i_{\rm sky},\Omega_{\rm sky})$. The
rotation between the disk and the sky frames requires, in addition, the
orientation of the disk plane, which is not known a priori: it enters the
joint fit discussed in Sect.~\ref{sec:joint_strategy} as a model parameter, so
for any assumed disk orientation the transcription is fully determined. The
same projection recovers the sky-frame direction of the spin from
$(\beta,\lambda_{\rm sky})$, where $\lambda_{\rm sky}$ is the spin azimuth
measured from the sky-plane ascending node.

The LT nodal precession, $\DO_{\mathrm{LT}}$, scales as $\rp^{-3/2}$, growing steeply as the pericenter shrinks.
Ignoring for the time being the different orientations, for S301 ($\rp = 280\,\rg$) and S2 ($\rp = 2800\,\rg$), the ratio of
LT signals is
\begin{equation}
  \frac{\DO_{\mathrm{LT}}^{\mathrm{S301}}}
       {\DO_{\mathrm{LT}}^{\mathrm{S2}}}
  =
  \left(\frac{\rp^{\mathrm{S2}}}{\rp^{\mathrm{S301}}}\right)^{\!\!3/2}
  \,
    \left(\frac{1+e_{\mathrm{S2}}}{1+e_{\mathrm{S301}}}\right)^{\!\!3/2}
  \approx 29,
  \label{eq:LT_ratio}
\end{equation}
while the ratio of disk precession rates is of order unity {for any disk extending beyond S2's pericenter}
(Sect.~\ref{sec:numerical}).  Using
Eqs.~(\ref{eq:embedded}) and (\ref{eq:DOLT1}), the disk-to-LT
confusion ratio in the  embedded regime for the $\alpha=0$ profile is
\begin{equation}
  \frac{\DO_{\mathrm{disk}}}{\DO_{\mathrm{LT}}}
  \sim \frac{\sin i\,\cos i}{2\chi}\frac{\Mdisk}{\MBH}
  \frac{1}{(1-e)^{1/2}}\left(\frac{\rp}{\rg}\right)^{\!\!3/2},
  \label{eq:ratio}
\end{equation}
which grows steeply with $\rp$, confirming that stars with smaller
pericenters are far less disk-confused.
The $(1-e)^{-1/2}$ factor is specific to $\alpha=0$; for the Mestel
profile $\alpha=-1$ the eccentricity dependence nearly cancels
(Sect.~\ref{sec:disk}) and this factor is absent -- for S301 the difference
amounts to a factor of $1/\sqrt{1-e}\simeq7.6$. For $\alpha=-1$ we therefore
quote the numerically computed ratios (Sect.~\ref{sec:numerical},
Table~\ref{tab:numerical}).
We assume hereafter $\chi = 1$
(maximal spin).  For $i = 30^\circ$, $\omega = 90^\circ$, $\alpha = -1$,
$\Mdisk = 10^3\,M_\odot$, and $\MBH = 4.3\times10^6\,M_\odot$:
\begin{align}
  \left|\frac{\DO_{\mathrm{disk}}}{\DO_{\mathrm{LT}}}
  \right|_{\mathrm{S301}} &\sim 1.5, &
  \left|\frac{\DO_{\mathrm{disk}}}{\DO_{\mathrm{LT}}}
  \right|_{\mathrm{S2}}   &\sim 14.
  \label{eq:confusion_ratios}
\end{align}
While S301 is $\sim 10$ times less confused than S2, the disk signal
remains comparable to LT for $\Mdisk \sim 10^3\,M_\odot$, motivating
the joint strategy of Sect.~\ref{sec:joint_strategy}.

For a disk compact enough to fit inside S301's own pericenter,
$r_{\rm disk} < r_{\rm p}^{\mathrm{S301}}$, the confusion ratio is instead
governed by the exterior-quadrupole regime (Eq.~\ref{eq:exterior}).
Combining it with Eq.~(\ref{eq:DOLT1}),
\begin{equation}
\left|\frac{\Delta\Omega_{\rm disk}}{\Delta\Omega_{\rm LT}}\right|
\simeq \frac{3k(\alpha)}{4\sqrt{2}\,\chi}\frac{M_{\rm disk}}{M_{\rm BH}}
\frac{r_{\rm disk}^{2}}{r_{\rm g}^{3/2}\,\left(r_{\rm p}^{\mathrm{S301}}\right)^{1/2}}
\sin i\,\cos i \lesssim 0.05,
\label{eq:compact_ratio}
\end{equation}
for $\alpha=-1$, $M_{\rm disk}=10^3\,M_\odot$, and $r_{\rm disk}\le
r_{\rm p}^{\mathrm{S301}}$, decreasing further as $r_{\rm disk}^{2}$. 
{In this regime, however, $M_{\rm disk}$ is not observationally
bounded by $10^{3}\,M_\odot$: a distribution buried inside S2's
pericenter affects S2 only through its monopole, which is absorbed into
the fitted black-hole mass, and through its quadrupole,
$Q\sim M_{\rm disk}r_{\rm disk}^{2}$. The S2 nodal bound
(Sect.~\ref{sec:disk}) constrains only the combination
$Q\cos i_{\rm S2}$, and permits masses as large as
$\sim\!10^{5}\,M_\odot$ for $r_{\rm disk}\approx r_{\rm p}^{\rm S301}$,
for which the ratio (Eq.~\ref{eq:compact_ratio}) reaches of order ten. Such a
configuration is, however, ruled out by astrophysical arguments  (see Sect.~\ref{sec:disk}). }

Because frame dragging precesses the orbital plane about the spin
axis, the nodal precession is accompanied by a companion
inclination drift,
\begin{equation}
  \Delta i_{\mathrm{LT}}
  = \frac{4\pi\chi\,\sin\beta\cos\lambda}{(1-e^2)^{3/2}}
    \left(\frac{\rg}{\rorb}\right)^{\!\!3/2} .
  \label{eq:DiLT}
\end{equation}
No kinematic projection factor appears in the inclination rate, since the
inclination is not measured within the reference plane
(Fig.~\ref{fig:decomp}).
In a
vector description \citep{BarkerOConnell1975, DamourSchafer1988}, frame
dragging rigidly precesses the orbital plane about the spin,
$\dot{\hat{\mathbf{L}}}\propto\chi\,\hat{\mathbf{S}}\times\hat{\mathbf{L}}$:
the plane precession (the nodal and inclination drifts) constrains the spin
component perpendicular to the orbital angular momentum, $\chi\sin\beta$, and
its azimuth $\lambda$, while the LT apsidal term constrains the parallel
component, $\chi\cos\beta$. Together they determine the full spin vector.
The exact orbital-element rates depend on the reference plane and
angular conventions \citep{will2008}. Eqs.~(\ref{eq:DOLT1}) and
(\ref{eq:DiLT}) are the rates referred to the disk plane. 
Both plane-precession components share the nodal timescale. The
apsidal advance, by contrast, must be disentangled from the far larger Schwarzschild precession, making it a longer-term measurement
(Sect.~\ref{sec:apsidal}).

\section{Numerical validation}
\label{sec:numerical}
The nodal precession per orbit was computed numerically for both stars
using the secular torque formula (Eq.~\ref{eq:dOdt}).  The
orbit-averaged disk potential is defined as
\begin{equation}
  \langle\Phi_{\rm disk}\rangle
  = -G \int_{r_{\rm in}}^{r_{\rm out}}
    2\pi\Sigma(r')\,r'\,\langle\Delta^{-1}\rangle_t\,\dd r',
  \label{eq:phi_avg}
\end{equation}
where
\begin{equation}
  \langle\Delta^{-1}\rangle_t
  = \frac{\oint \Delta^{-1}(r,r',f)\,r^2\,\dd f}
         {\oint r^2\,\dd f}
  \label{eq:inv_delta}
\end{equation}
is the Kepler-time-weighted, azimuthally averaged inverse distance
between the star at orbital phase $f$ and a disk ring at radius $r'$,
with $\Delta^2 = |\mathbf{r} - \mathbf{r}'|^2 + H^2$ softened by the
disk scale height $H = 0.01\,r'$ to regularize the divergence at the
disk-crossing points.
The derivative
$\partial\langle\Phi_{\rm disk}\rangle/\partial i$ was evaluated by a
{centered } finite difference with step $\delta i = 0.3^\circ$.
The orbital phase average was calculated using $r^2$  Kepler weighting.
Both  stars  share the same disk
geometry and are scanned over the same grid of inclinations, $i$, and
arguments of pericenter, $\omega$, relative to the disk plane. We considered several disk configurations (different radial density power-law distributions and different inner and outer edges).

\begin{figure}[t]
\centering
\includegraphics[width=\hsize]{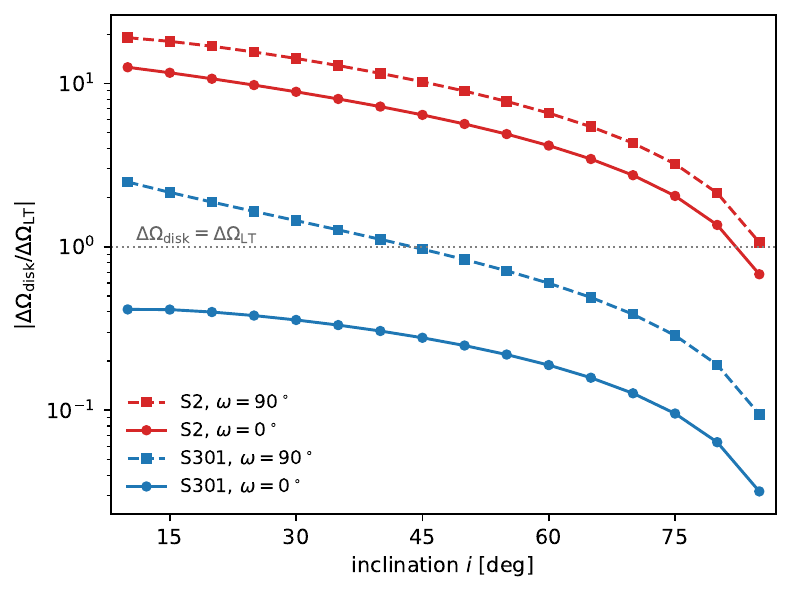}
\caption{Disk-to-LT confusion ratio, $|\DO_{\rm disk}/\DO_{\rm LT}|$, as a
function of the orbit--disk inclination, $i$, for S301 (blue) and S2 (red)
and for $\omega=0^\circ$ (solid circles) and $\omega=90^\circ$ (dashed
squares), computed for the $\alpha=-1$ disk of Table~\ref{tab:numerical}.
The dotted line marks $\DO_{\rm disk}=\DO_{\rm LT}$. S301 lies below unity
at all geometries except $\omega=90^\circ$ with $i\lesssim45^\circ$,
whereas S2 is disk-dominated almost everywhere -- the $\sim10\times$
vertical offset between the two stars is the pericenter lever arm of
Eq.~(\ref{eq:LT_ratio}). The ratios vanish at $i=90^\circ$, where the disk
torque ($\propto\cos i$) goes to zero.}
\label{fig:confusion}
\end{figure}

Table~\ref{tab:numerical} and Fig.~\ref{fig:confusion} present the full numerical results for both
stars with an $\alpha = -1$ disk and
$r_{\mathrm{disk}} = r_{\mathrm{a}}^{\mathrm{S301}}$.
The ratio
$|\DO_{\mathrm{disk}}^{\mathrm{S2}}/\DO_{\mathrm{disk}}^{\mathrm{S301}}|$
ranges from
$\sim 0.26$ to $\sim 1.0$
across all inclinations and
arguments of pericenter, confirming comparable disk torques (within a
factor of $\lesssim 4$ of unity) in contrast
to the factor of $\sim 30$ separating the LT signals.
As expected, the
individual confusion ratios in Table~\ref{tab:numerical} agree with
Eq.~(\ref{eq:confusion_ratios}), of which they are the numerical basis, and
are consistent, to order of magnitude, with the analytic scaling of
Eq.~(\ref{eq:ratio}) once the $\alpha=0$ eccentricity factor is removed.
Similar results
are obtained for other values of $\alpha$ and $r_{\mathrm{disk}}$, with
the disk effect decreasing, as expected, with decreasing $\alpha$: a
steeper profile places less mass at the large radii that dominate the
secular torque.

The variation with $\omega$ reflects the geometry of the
orbit relative to the disk plane.
The node-crossing radii are
\begin{equation}
  r_{\mathrm{asc}} = \frac{p}{1+e\cos\omega},
  \qquad
  r_{\mathrm{des}} = \frac{p}{1-e\cos\omega},
  \label{eq:crossing_radii}
\end{equation}
where $p = \rorb(1-e^2)$.
The nodal torque vanishes
at the crossings themselves -- the elementary torque about the node line
carries the factor $\sin(\omega+f)$, which is zero where the star lies in the
disk plane -- and the secular torque accumulates instead around the
antinodes, where the star's elevation above the plane is maximal. For
$\omega = 90^\circ$ the antinodes coincide with pericenter and apocenter:
the star is then maximally elevated at apocenter, where it spends the vast
majority of its orbital period, and the time-integrated torque is strongly
amplified. For $\omega = 0^\circ$ the antinodes lie at $r = p$, with a much
smaller elevation lever arm and residence time. This accounts for the factor of
$\sim 3$-$6$ variation between $\omega = 0^\circ$ and $\omega = 90^\circ$ in
Table~\ref{tab:numerical}.
The enhancement is
stronger for S301, whose higher eccentricity concentrates more of the orbital
period near apocenter; the ratio of disk torques therefore drops to
$\sim 0.3$ at $\omega = 90^\circ$, while remaining of order unity across all
geometries.
\begin{table}
\caption{Ratio $|\Delta\Omega_{\rm disk}/\Delta\Omega_{\rm LT}|$ and disk torque comparability.}
\label{tab:numerical}
\centering
\begin{tabular}{ccccc}
\hline\hline
Inclination & Arg. of pericenter & $\left|\frac{\Delta\Omega_{\rm disk}}{\Delta\Omega_{\rm LT}}\right|$ & $\left|\frac{\Delta\Omega_{\rm disk}}{\Delta\Omega_{\rm LT}}\right|$ & $\left|\frac{\Delta\Omega_{\rm disk}^{\mathrm{S2}}}{\Delta\Omega_{\rm disk}^{\mathrm{S301}}}\right|$ \\
($i$) & ($\omega$) & S301 & S2 & \\
\hline
$10^\circ$ & $0^\circ$  & 0.41  & 13  & 1.0 \\
$10^\circ$ & $90^\circ$ & 2.5   & 19    & 0.26 \\
$30^\circ$ & $0^\circ$  & 0.36  & 8.9   & 0.85 \\
$30^\circ$ & $90^\circ$ & 1.4  & 14    & 0.34 \\
$60^\circ$ & $0^\circ$  & 0.19  & 4.2   & 0.75 \\
$60^\circ$ & $90^\circ$ & 0.61  & 6.6   & 0.37 \\
\hline
\end{tabular}
\tablefoot{Computed numerically for an $\alpha=-1$ power-law disk layout with $M_{\rm disk} = 10^3 M_\odot$ and an outer disk edge at S301's apocenter ($r_{\rm disk} = r_a^{\mathrm{S301}}$). Reference LT values adopt maximal black hole spin $\chi=1$, yielding $\Delta\Omega_{\rm LT}^{\mathrm{S301}} = 9.6 \times 10^{-4}$ rad/orbit and $\Delta\Omega_{\rm LT}^{\mathrm{S2}} = 3.3 \times 10^{-5}$ rad/orbit. Both nodal rates are referred to the disk plane: the disk rates follow
from Eq.~\ref{eq:dOdt}, while the LT rates apply the kinematic projection
factor $1/\sin i$ (Eq.~\ref{eq:DOLT1}) to the reference values quoted above,
which are the frame-invariant plane-precession magnitudes.}
\end{table}

To verify that a disk produces the maximal Newtonian nodal precession for
a given radial mass profile, we repeated the calculation for oblate
spheroidal mass distributions with the same power-law density
$\rho(s) \propto s^\alpha$, where $s^2 = R^2 + z^2/q^2$ is the
ellipsoidal coordinate and $q$ is the axis ratio ($q=1$ is a
sphere, $q\to 0$ approaches a razor-thin disk).  The mass was kept fixed
at {$\Menc = 10^3\,M_\odot$} for all $q$.  The torque integral was
evaluated by computing the analytical gradient $\partial(1/\Delta)/\partial i$
at each orbital phase integrating over the meridional angle $\psi$ with  midpoint-rule  and integrating over the ellipsoidal shell
coordinate $s$ with  Gauss-Legendre quadrature.
As $q$ decreases toward the { razor-thin} disk limit the nodal
precession increases monotonically and converges to the disk result\footnote{{Note that the $q\to0$ limit of the density
$\rho(s)\propto s^{\alpha}$ corresponds to a razor-thin disk with
surface density $\Sigma\propto r^{\alpha+1}$.}}.
In all cases the disk ($q \to 0$) produces the largest nodal precession
for a given enclosed mass and radial profile, confirming that the disk
results in Table~\ref{tab:numerical} represent upper limits on the
Newtonian confusion.
\section{A joint strategy}
\label{sec:joint_strategy}
The apocenter-dominance of the secular disk torque, $\Delta\Omega_{\text{disk}}$, is the basis of the joint stellar approach. The total observed nodal precession rates of S301 and S2 can be written as a system of coupled equations:
\begin{equation}
\Delta\Omega_{\text{obs}}^{\mathrm{S301}} = \Delta\Omega_{\text{LT}}^{\mathrm{S301}}(\chi, \beta^{\mathrm{S301}}, \lambda^{\mathrm{S301}})
+ \Delta\Omega_{\text{disk}}^{\mathrm{S301}}(M_{\text{disk}}, i^{\mathrm{S301}}, \omega^{\mathrm{S301}}), 
\end{equation}
\begin{equation}
\Delta\Omega_{\text{obs}}^{\mathrm{S2}}= \Delta\Omega_{\text{LT}}^{\mathrm{S2}}(\chi, \beta^{\mathrm{S2}}, \lambda^{\mathrm{S2}}) 
+ \Delta\Omega_{\text{disk}}^{\mathrm{S2}}(M_{\text{disk}}, i^{\mathrm{S2}}, \omega^{\mathrm{S2}}).
\end{equation}
Analogous equations hold for S55 and S38; we write them for S2 alone for
clarity, and the joint fit of all reference stars is discussed below.

Because S2 and S301 share comparable apocenters ($r_{\rm a}^{\mathrm{S2}} \approx 45400\,\rg$ versus $r_{\rm a}^{\mathrm{S301}} \approx 31700\,\rg$), they track the same underlying secular torque regime 
{for any disk extending beyond S2's pericenter, regardless of its profile; more compact configurations are excluded in Sect. \ref{sec:disk}.}
Conversely, because the frame-dragging component scales steeply with pericenter distance, $\rp^{-3/2}$, their LT signals differ by a factor of $\approx 30$. Subtracting the observed signals may enable us to isolate the frame-dragging contribution
as the   Newtonian background largely cancels out and robustly constrain the absolute mass profile of the disk \citep[e.g.,][]{gravity2024}. However, the unknown disk orientation and black hole spin orientations complicate this naive  cancellation approach.

Counting parameters makes the solvability of the joint problem
explicit. The disk model has five parameters: the total mass, outer
radius, and power-law exponent, plus two orientation angles; frame
dragging adds three more: the spin magnitude $\chi$ and two
orientation angles. Against these eight unknowns, each star supplies
up to three secular rates --- the two plane-precession components
($\dot\Omega$ and the inclination drift, $\dot\imath$) and the apsidal
rate, whose information resides in the residual after the
Schwarzschild subtraction. The apocenter-matched reference stars S2,
S55, and S38 alone provide six plane-precession rates as compare with the five
disk parameters, so the Newtonian foreground is over-determined without
invoking S301. S301's  plane precession then brings  two
in-plane spin components, its apsidal residual the third
(Sect.~6).  S29's wider orbit adds independent leverage on the
radial mass profile. The joint problem is thus over-determined. This
counting establishes identifiability rather than precision: the
weight of each rate in the fit is set by how well it can be measured.

The analytic estimates and the numerical results of
Table~\ref{tab:numerical} assume the same inclination, $i$, and argument
of pericenter, $\omega$, for both stars relative to the disk plane. In
reality S2 and S301 have different orientations, and the disk plane
itself is an unobserved free parameter, so the true $(i,\omega)$ of
each star relative to the disk differ. The disk torques therefore
carry different geometric factors, and their ratio can fluctuate
between
$\sim 0.26$ and $\sim 1.0$ (Table~\ref{tab:numerical}) depending on
the node-crossing tracks. This residual geometric mismatch must be
fitted jointly within the global orbital model \citep{iorio2011}.

The modeling is most demanding for orbits nearly coplanar with the disk: as shown in  Fig.~\ref{fig:confusion}, for $i=10^\circ$ and $\omega=90^\circ$ S301's disk-to-LT confusion ratio reaches
$\simeq 2.5$, i.e., the quadrupole background exceeds the relativistic signal by a factor of a few.
However, because the disk torques on both stars are comparable while their LT rates differ by the factor $\Delta\Omega_{\text{LT}}^{\mathrm{S301}} / \Delta\Omega_{\text{LT}}^{\mathrm{S2}} \approx 30$, the joint differential signal, rather than S301's precession taken alone, remains usable even in this case.

The strength of the plane-precession signal depends on the spin
orientation: both $\DO_{\mathrm{LT}}\propto\chi\sin\beta\sin\lambda$ and the
inclination drift $\Delta i_{\mathrm{LT}}\propto\chi\sin\beta\cos\lambda$
scale with the in-plane projection $\chi\sin\beta$.
If the spin lies close to S301's orbital axis ($\sin\beta$ small), this in-plane
signal is weak and the spin can be inferred instead through the apsidal
precession, {$\Delta\widetilde\varpi_{\mathrm{LT}}\propto\chi\cos\beta$}
 - the orthogonal (see Sect.~\ref{sec:apsidal}). The relative
scale of the competing nodal contributions for S301 is summarized in
Table~\ref{tab:nodal_budget}.
\begin{table}
    \caption{Nodal precession budget for S301 (rad/orbit).}
    \label{tab:nodal_budget}
    \centering
    \begin{tabular}{ccc}
        \hline
        \hline
        Component & Nature & Strategy \\
        \hline
        Lense-Thirring  & Frame- & Primary signal [3]. \\
        $\Delta\Omega_{\text{LT}} \sim 9.6 \times 10^{-4}$ & Dragging [2] &   \\
        \hline
        Newtonian Disk & Axisymmetric & Other stars [3]; \\
        $\Delta\Omega_{\text{disk}} \sim 2.8 \times 10^{-3}$ & Quadrupole & $\Omega$-rate tracking [4,5]. \\
        \hline
        Schwarzschild & Pure GR & Zero first order; \\
        $\Delta\Omega_{\text{Schw}} = 0$ & Monopole [1] & acts as orbital clock [1]. \\
        \hline
    \end{tabular}
    \tablefoot{Fiducial values for $M_{\rm disk}=10^3\,M_\odot$, $\alpha=-1$,
$r_{\rm disk}=r_{\rm a}^{\mathrm{S301}}$; the quoted disk value is for $i=30^\circ$, $\omega=90^\circ$, the geometry of Eq.~\ref{eq:confusion_ratios}. The disk
term varies with the orbit--disk geometry from $\sim2\times10^{-4}$ to
$\sim1.4\times10^{-2}$ rad\,orbit$^{-1}$ across the configurations of
Table~\ref{tab:numerical}; the LT value assumes $\chi=1$ and favorable orientation.
The LT entry is the frame-invariant plane-precession magnitude; referred
to the disk plane, the LT nodal rate is larger by the kinematic factor
$1/\sin i$ (Sect.~\ref{sec:LT_contrast}).
    References: [1] \citet{gravity2020}, [2] \citet{lense1918}, [3] \citet{merritt2010}, [4] \citet{BrouwerClemence1961}, [5] \citet{MurrayDermott1999}.}
\end{table}

Several stars with comparable apocenters around Sgr~A*
provide an observational test to separate the Newtonian cluster background
from the LT precession. These stars have already been used to limit the
monopole of the mass distribution around Sgr~A* \citep{gravity2022,gravity2024};
their orbital parameters are compiled in Table~\ref{tab:combined_apocenters_rg}
and illustrated in Fig.~\ref{fig:orbits}. Although S301 has an exceptionally
close pericenter, its apocenter ($\ra^{\mathrm{S301}} \approx 31{,}700\,\rg$)
is comparable to those of S55, S2, and S38
($35{,}000\,\rg \lesssim \ra \lesssim 50{,}000\,\rg$). {Because the secular
Newtonian torque is governed by the ratio $\ra/r_{\rm disk}$
(Sect.~\ref{sec:disk}), these three stars sample essentially the same
extended-mass environment as S301, despite pericenters that are
much larger and LT signals that are correspondingly negligible.}
{S29, whose apocenter ($\approx 1.5\times10^5\,\rg$) is several times
larger (Fig.~\ref{fig:orbits}, left), instead probes the mass distribution
at larger radii and anchors its radial profile.}

The mutually misaligned orbital planes of the reference stars
are essential: because the disk torque on each star
depends on its own orientation relative to the disk, only a set of
differently oriented orbits can determine the disk plane in three
dimensions, and all four reference stars enter the joint fit.
Monitoring of
these additional orbits isolates the global Newtonian background: by fitting
their collective secular precession rates, a direct measurement - or a
rigorous upper limit - on the distributed mass profile can be established,
absorbing the disk degeneracy and isolating the pure relativistic LT
precession signal of S301.

{Both S301 orbit solutions (A and B; Fig.~\ref{fig:orbits}) place
the star in the same apocenter-matched regime as S2, so the
disk-cancellation strategy is insensitive to the orbit ambiguity. The
spin-projection factors $\sin\beta\sin\lambda$, however, differ between A
and B and must be resolved in the joint fit (Sect.~\ref{sec:apsidal}).}
\begin{table}
    \caption{Orbital parameters and derived apocenters in gravitational
    radii ($\rg$).}
    \label{tab:combined_apocenters_rg}
    \centering
    \begin{tabular}{lccc}
        \hline
        \hline
        Star & Semimajor  & Eccentricity ($e$) & Apocenter ($r_a$) \\
             & Axis ($a$) [$\rg$] & & [$\rg$] \\
        \hline
        \textbf{S301} & 16,010 & 0.9825 & 31,740 \\
        \textbf{S55}  & 20,334 & 0.7267 & 35,110 \\
        \textbf{S2}   & {24,100} & 0.8844 & {45,400} \\
        \textbf{S38}  & 27,763 & 0.8145 & 50,386 \\
        \textbf{S29}  & 77,421 & 0.9693 & 152,465 \\
        \hline
    \end{tabular}
    \tablefoot{The newly discovered star S301 is combined with the core
    empirical tracking baselines from \citet{gravity2022}.
    Distances adopt $\MBH = 4.297\times10^{6}\,M_\odot$ and
    $R_0 = 8.28$~kpc ($1'' \approx 195{,}350\,\rg$); apocenters are
    $\ra = \rorb(1+e)$. Parameters for S301 are from \citet{abdeldayem2026}.}
\end{table}
\section{Apsidal precession and spin vector resolution}
\label{sec:apsidal}
The joint nodal strategy of Sect.~\ref{sec:joint_strategy} returns the
nodal LT rate $\DO_{\mathrm{LT}}\propto\chi\sin\beta\sin\lambda/\sin i$, one
component of the spin. The inclination drift
(Sect.~\ref{sec:LT_contrast}) supplies the second in-plane component,
$\chi\sin\beta\cos\lambda$; together the two plane-precession rates fix the
in-plane projection of the spin, $\chi\sin\beta$, and its orientation
$\lambda$. Because the orbital plane is unaffected by the Schwarzschild
precession, these two rates compete only with the disk (Sect.~\ref{sec:joint_strategy}).
The remaining component, the projection $\chi\cos\beta$ along the orbital
axis, is carried only by the apsidal precession, to which we now turn.

After subtracting the dominant Schwarzschild general relativistic advance, $\Delta\omega_{\text{Schw}} = 6\pi \rg / [\rorb(1-e^2)]$, the remaining post-Newtonian apsidal shift is governed by a distinct angular dependence. Because the pericenter argument is measured from the ascending node, which itself precesses, the raw rate $\Dw_{\text{LT}}$ contains the kinematic term $-\cos i\,\DO_{\text{LT}}$ and hence inherits a dependence on the transverse spin components and on the inclination. The combination {$\widetilde \varpi \equiv \omega + \cos i\,\Omega$ } removes this nodal shift, and it is this node-corrected apsidal advance that carries the parallel spin component:
{\begin{equation}
\Delta\widetilde\varpi_{\rm LT} = \Delta\omega_{\rm LT,sky}
  + \cos i_{\rm sky}\,\Delta\Omega_{\rm LT,sky}
  \propto \chi\cos\beta .
\label{eq:apsidal}
\end{equation}
The node-corrected apsidal combination $\widetilde\varpi$, introduced in
Sect.~\ref{sec:LT_contrast}, is written here in the sky frame, where all the
elements are directly measured.}
Because the nodal shift scales as
$\DO_{\text{LT,sky}} \propto \chi \sin\beta
\sin\lambda_{\rm sky}/\sin i_{\rm sky}$, taking the ratio of these two frame-dragging signatures for a single star completely cancels out the spin magnitude, $\chi$, isolating the pure orientation geometry of the system:
{\begin{equation}
\frac{\DO_{\text{LT,sky}}}{\Delta\widetilde \varpi_{\text{LT}}}
\propto \frac{\sin\beta \sin\lambda_{\rm sky}}{\sin i_{\rm sky}\,\cos\beta}
= \frac{\tan\beta \sin\lambda_{\rm sky}}{\sin i_{\rm sky}}.
\end{equation}}

Recovering $\chi\cos\beta$ from the apsidal advance is a longer-term
measurement. The LT apsidal rate, $\Dw_{\mathrm{LT}}\sim2\times10^{-3}$~rad~
orbit$^{-1}$, is an order of magnitude below the Schwarzschild advance
$\Dw_{\mathrm{Schw}}\sim3.4\times10^{-2}$~rad~orbit$^{-1}$, and is separated
from it only as the pericenter sweeps through many orbits
(Sect.~\ref{sec:time}); the disk apsidal term must likewise be modeled with
the reference stars. At S301's pericenter the second order post-Newtonain (2PN)  correction to the monopole
apsidal advance is 
$\Dw_{\mathrm{2PN}} \simeq
[(18+e^2)/4]\,(\rg/p)\,\Dw_{\mathrm{Schw}} \approx
2.9\times10^{-4}$~rad~orbit$^{-1}$, a factor of $\simeq 7$ below
$\Dw_{\mathrm{LT}}$ and $\simeq 4$ times larger than the spin-quadrupole term
below; being a parameter-free function of $\MBH$ and the
orbit, it is absorbed into the Schwarzschild subtraction, which must
therefore be carried out to 2PN accuracy. The spin-quadrupole (no-hair)
terms, by contrast, are smaller than the LT rates by
$\sim\chi(\rg/p)^{1/2}\approx 4\%$ and are negligible here. The apsidal precession therefore completes the
three-dimensional spin vector, adding $\chi\cos\beta$ to the in-plane
projection already fixed by the plane precession, but on a timescale of
several orbits, i.e.,\ a few decades, rather than the one or two orbits
needed for the plane precession.

Isolating this spin signature requires disentangling it from the broader
background. Through its Newtonian monopole, the apsidal precession constrains
the spherically averaged enclosed mass $M(<\rorb)$ at a star's
semimajor axis, independently of the disk geometry. For S2 this established the fiducial
bound {$\Menc \lesssim 10^3\,M_\odot$} \citep{gravity2024}. The other
apocenter-matched stars extend the constraint over a range of radii: S55,
S2, and S38 sample $\rorb \approx 20{,}000$-$28{,}000\,\rg$ and S29 reaches
$\sim 77{,}000\,\rg$ (Table~\ref{tab:combined_apocenters_rg}) - so that
S301, at the innermost $\rorb^{\mathrm{S301}} \approx 16{,}000\,\rg$, adds
the deepest point of an empirically determined profile $M(<\rorb)$ rather
than relying on a single anchor. The same set of stars fixes the disk model,
its mass, plane, and flattening (Sect.~\ref{sec:joint_strategy}), which is needed to predict and remove the Newtonian quadrupole term in S301's
apsidal budget.

However, the primary dynamical challenge for S301's periapsis shift is one of
scale. As shown in Table~\ref{tab:apsidal_budget}, the LT signal
lies between a massive Schwarzschild background and a competitive Newtonian
disk perturbation.
\begin{table}
    \caption{Apsidal precession budget for S301 (rad/orbit).}
    \label{tab:apsidal_budget}
    \centering
    \begin{tabular}{ccc}
        \hline
        \hline
        Component & Nature & Strategy \\
        \hline
        Schwarzschild  & Pure GR & Dominant [1]; \\
        $\Delta\omega_{\text{Schw}} \sim 3.4 \times 10^{-2}$ & Monopole & easily subtracted. \\
        \hline
        Lense-Thirring  & Frame- & Constant offset \\
        $\Delta\omega_{\text{LT}} \sim 2.0 \times 10^{-3}$ & Dragging &  via $\chi \cos\beta$ [2,3]. \\
        \hline
        Newtonian Disk & Axisymmetric & Time-varying \\
        $\Delta\omega_{\text{disk}} \sim 7.0 \times 10^{-4}$ & Quadrupole & via $\omega$-advance [4,5]. \\
        \hline
    \end{tabular}
    \tablefoot{References: [1] \citet{gravity2020}, [2] \citet{lense1918}, [3] \citet{merritt2010}, [4] \citet{BrouwerClemence1961}, [5] \citet{MurrayDermott1999}.}
\end{table}

The large Schwarzschild advance, $\Dw_{\mathrm{Schw}} \approx
1\fdg95$~orbit$^{-1}= 3.4 \times 10^{-2}$~rad orbit$^{-1}$, can serve as an intrinsic clock,
rotating the argument of pericenter $\omega$ appreciably over just a few
orbital periods. Because the disk quadrupole term $\Dw_{\mathrm{disk}}(\omega)$
depends on the orbit's orientation relative to the disk whereas
$\Dw_{\mathrm{LT}}$ does not, the time-varying part of the apsidal signal is a
pure disk diagnostic, aiding its separation from the constant LT term
(Sect.~\ref{sec:time}).
\section{Granularity of the Newtonian background}
\label{sec:granularity}
So far  we have treated
the Newtonian background as a smooth, axisymmetric disk, an
idealization that maximizes the nodal precession for a given enclosed
mass (Sect.~\ref{sec:disk}) and hence yields a conservative upper
bound on the confusion. Granularity of the perturber population is primarily a caveat on the
accuracy of the Newtonian subtraction: it degrades the precision of the
strategy rather than its viability, converting part of the subtraction
from a deterministic correction into a stochastic one.  \citet{Bordoni+2025} simulated S2's
orbit through a cluster of $N$ discrete bodies (motivated by mass
segregation, which concentrates stellar-mass black holes in the
innermost region), showing that for a fixed enclosed mass
$\Menc=1000\,M_\odot$ the orbital-plane precession and its
orbit-to-orbit scatter grow with the individual perturber mass: while the effect is small for solar mass stars or stellar remnants,
$100\,M_\odot$ objects drive a  precession  comparable
to the smooth-disk secular torque itself
(Table~\ref{tab:nodal_budget}). This can bias a smooth-model fit of the
enclosed mass by up to a factor of $\sim6$, and in a non-negligible
fraction of realizations even mimic a net prograde signal.

Such granularity may set a stochastic floor beyond
the $\sim 3$--$4$ geometric factor of Table~\ref{tab:numerical}: the
reference-star calibration of the nodal budget
(Sect.~\ref{sec:joint_strategy}), the $\Dw_{\mathrm{disk}}$ term of
the apsidal budget (Sect.~\ref{sec:apsidal}), and the time-variation
diagnostic (Sect.~\ref{sec:time}) all acquire a residual that is
neither constant like LT nor cleanly periodic in $\omega$ like the
smooth-disk term: even a perfectly known total enclosed mass need
not translate into a fixed, repeatable Newtonian precession. Averaging
over S301's many orbits during its rapid Schwarzschild sweep, and over the ensemble of reference stars, suppresses this floor but does not
fully remove it. A distinct systematic is the Brownian recoil of
Sgr~A* itself - displacements of up to $6\,\mu\mathrm{as}$ 
{in the simulations of \citet{Bordoni+2025}, 
for $100\,M_\odot$ perturbers
 with the cusp-like radial distribution adopted there; the amplitude scales with the spatial extent of the perturber population - }
 a shift of the common astrometric
reference point that enters a joint multi-star fit as correlated noise
rather than a per-star torque.

Interestingly, \citet{Bordoni+2025} reach the same qualitative
conclusion that motivated this work: the LT precession remains
steeply peaked at the pericenter regardless of whether the Newtonian
background is smooth or granular, preserving, in principle, the
separation between spin and perturbations \citep{ZhangIorio2017}.
Quantifying this separation for S301 specifically, however, requires
extending their granular framework to a pericenter nearly two orders
of magnitude tighter than S2's, where the relevant perturber count and
the validity of a fixed-cusp approximation both remain to be
established; we leave this to future work.

{This behavior reflects the finite-$N$ fluctuations of the
perturber distribution: the mean precession is set by the smooth
component of the potential and is independent of the perturber mass
($m$) at fixed $M_{\rm enc}$, while the torques of $N$ randomly oriented
perturbers add incoherently, producing a residual that scales as
$\sqrt{N}\,m=\sqrt{m\,M_{\rm enc}}$, the vector-resonant-relaxation
scaling \citep{rauch1996}. The orbital-plane change quoted from the
simulations is a positive-definite magnitude, and therefore grows with
$m$ even though the ensemble-averaged precession does not.}

\section{The time-variation discriminant}
\label{sec:time}
As $\omega$ advances due to Schwarzschild precession, the node-crossing
radii, Eq.~\ref{eq:crossing_radii}, change, and the disk contributions
$\DO_{\mathrm{disk}}(\omega)$ and $\Dw_{\mathrm{disk}}(\omega)$ vary
from orbit to orbit, while the LT rates and the monopole apsidal term
remain independent of $\omega$.  The time-varying component of the
observed precession is therefore a pure disk signature, and the constant
component is a mixture of LT and the $\omega$-averaged disk.

For S301, as $\omega$ advances the node-crossing radii
(Eq.~\ref{eq:crossing_radii}) migrate from
$(r_{\mathrm{asc}},r_{\mathrm{des}}) = (\rp,\ra) \simeq
(280,\,31700)\,\rg$ at $\omega = 0$ to
$r_{\mathrm{asc}} = r_{\mathrm{des}} = p \equiv \rorb(1-e^2) = 555\,\rg$ at
$\omega = 90^\circ$, while the antinodes, at which the secular torque is
applied (Sect.~\ref{sec:numerical}), migrate correspondingly from $r = p$
to pericenter and apocenter, driving a factor of $\sim 3$-$6$ variation in
$|\DO_{\mathrm{disk}}/\DO_{\mathrm{LT}}|$ (Table~\ref{tab:numerical}).
The rapid Schwarzschild apsidal advance of S301
($\Dw_{\mathrm{Schw}} \approx 1\fdg95$~orbit$^{-1}$, versus
$0\fdg20$~orbit$^{-1}$ for S2) means this variation plays out over
a few dozen orbital periods, making it  observationally accessible but over a rather long time period.
\section{Conclusions}
\label{sec:conclusions}
We have shown that while matching apocenters calibrate the Newtonian background, a mismatched pericenter isolates the spin. As the secular disk torque, and hence
$\DO_{\mathrm{disk}}$, is set by the apocenter rather than the pericenter, this ratio
can be used to break the possible degeneracy between nodal precession of S301 due to LT and due to a nonspherical Newtonian mass distribution. Stars sharing S301's apocenter but with much larger
pericenters  specifically S2, but also  S55 and S38, feel comparable disk torques yet have negligible
LT precession. Their observed precession is therefore an almost
pure Newtonian signal that maps {$\Menc$} and the matter orientation  (extending the apsidal
mass bound of \citealt{gravity2020}), while the wider orbit of S29 anchors
the radial profile. With the mass distribution so calibrated, its contribution to S301's
precession can be modeled and removed, leaving the relativistic signal of S301 detectable. 
{This calibration is what promotes the idealized $\sigma_\chi < 0.2$ forecast of the discovery paper \citep{abdeldayem2026}, obtained for a pure Kerr metric, into a robust measurement in the presence of the Newtonian foreground.}

In fact, current data on these stars can already be used to set
interesting limits. {As shown in Sect.~\ref{sec:disk},} the
{non-detection of nodal precession for S2 bounds it} to
$\lesssim 3^{3/2}\,\Delta\Omega^{\rm S2}_{\rm LT}\approx
1.7\times10^{-4}$~rad~orbit$^{-1}$, already a direct observational
limit on the Newtonian torque acting on S2. This, in turn, implies
that, within a rough approximation, the quadrupole of the mass
distribution interior to $\sim r_{\rm a}^{\rm S2}$ is a factor of
$\sim3$ smaller than the maximal value adopted in our estimates (a
razor-thin disk with $M_{\rm disk}=10^{3}\,M_\odot$ {truncated at
$r_{\rm disk}=r_{\rm a}^{\rm S301}$}). Correspondingly, this reduces
the disk-to-LT confusion of S301 (Eq.~7).

The large pericenter ratio,
$(\rp^{\mathrm{S2}}/\rp^{\mathrm{S301}})^{3/2}\approx30$,
gives S301 a much larger relativistic signal than the reference stars, making the residual Newtonian contamination calibratable in a joint fit; and as S301's pericenter advances under its rapid
Schwarzschild precession, the Newtonian rates vary from orbit to orbit while the
LT rates stay constant, separating the time-varying disk signal from the
constant LT offset (Sect.~\ref{sec:time}).

These calibration and time-domain strategies assume a smooth Newtonian background. As discussed in Sect.~\ref{sec:granularity}, if the enclosed mass is dominated by a few massive perturbers, granularity adds a stochastic, orbit-to-orbit component that is neither constant like the LT terms nor periodic in $\omega$ like the smooth-disk signal, setting a residual noise floor that ensemble-averaging over the reference stars and over S301's many orbits can suppress but not fully remove. Encouragingly, the steep pericenter concentration of the LT precession survives in the granular case as well, so the separation advocated here remains viable, though quantifying it at S301's pericenter awaits extension of the granular simulations.

Once the Newtonian effect is removed, S301's plane precession, its nodal and
inclination drifts, {measures the spin component perpendicular to the orbital axis:} the in-plane
projection $\chi\sin\beta$ and its orientation $\lambda$. Because the orbital
plane carries no Schwarzschild precession, this is achievable within a few
orbits and is the near-term goal. The third component, $\chi\cos\beta$, is
carried by the apsidal advance; separating it from the dominant Schwarzschild
term requires the pericenter to sweep through many orbits, so it completes the
full spin vector only on a longer timescale. The principal residual
uncertainties are the twofold orbit solution for S301 (A and B;
Fig.~\ref{fig:orbits}) and the modeling of the subtracted Newtonian torque, including its granularity (Sect.~\ref{sec:granularity}). With
continued GRAVITY+ astrometry and ELT spectroscopy of the apocenter-matched
reference stars,  a measurement of both the magnitude and direction of the spin of Sgr~A* will become feasible.

\begin{acknowledgements}
This work was supported by the Simons Foundation SCEECS collaboration (MP-SCMPS-00001470) and by an ERC advanced grant MultiJets. DC and JC acknowledge the financial support from ANID-FONDECYT Regular 1251444.
The research of DC has been funded by the Alexander von Humboldt Foundation.
We are grateful to the anonymous referee for constructive comments. 

\end{acknowledgements}
\bibliographystyle{aa}

\end{document}